# Scanning tunneling microscopy and spectroscopy of twisted trilayer graphene

Wei-Jie Zuo, Jia-Bin Qiao, Dong-Lin Ma, Long-Jing Yin, Gan Sun, Jun-Yang Zhang, Li-Yang Guan, and Lin He*

Center for Advanced Quantum Studies, Department of Physics, Beijing Normal University, Beijing, 100875, People's Republic of China
*Email: helin@bnu.edu.cn

**Twist, as a simple and unique degree of freedom, could lead to enormous novel quantum phenomena in bilayer graphene. A small rotation angle introduces low-energy van Hove singularities (VHSs) approaching the Fermi level, which result in unusual correlated states in the bilayer graphene. It is reasonable to expect that the twist could also affect the electronic properties of few-layer graphene dramatically. However, such an issue has remained experimentally elusive. Here, by using scanning tunneling microscopy/spectroscopy (STM/STS), we systematically studied a twisted trilayer graphene (TTG) with two different small twist angles between adjacent layers. Two sets of VHSs originating from the two twist angles were observed in the TTG, indicating that the TTG could be simply regarded as a combination of two different twisted bilayer graphene. By using high-resolution STS, we observed split of the VHSs and directly imaged spatial symmetry breaking of electronic states around the VHSs. These results suggest that electron-electron interactions play an important role in affecting the electronic properties of graphene systems with low-energy VHSs.**

Twisted bilayer graphene (TBG), i.e., bilayer graphene with a rotation angle[1–23], exhibits a variety of emergent phenomena and properties, such as the Dirac fermions confined by non-Abelian gauge potentials[8,16], Fermi velocity renormalization[18,19], and novel correlated states[23–32], that can be simply tuned by a twist. Therefore, twist could be treated as a unique degree of freedom to tune the structures and electronic properties of graphene systems. Very recently, a small rotation angle was also introduced into multilayer graphene[33-37]. It was experimentally demonstrated that the twist dramatically changes the structures of multilayer graphene and, consequently, it was reasonable to expect that the twist may introduce novel electronic properties in the multilayer graphene. However, the effects of the twist on the electronic properties of multilayer graphene are much less explored. For example, an experimental study on electronic properties of twisted trilayer graphene (TTG), which has two different rotation angles between adjacent layers, is still lacking up to now.

In this paper, we systematically studied both the topographic structures and the low-energy electronic properties of the TTG by using scanning tunneling microscopy/spectroscopy (STM/STS). The morphology of the TTG was characterized by "double-moiré" patterns because of the existence of two different twist angles between adjacent layers. Two sets of low-energy van Hove singularities (VHSs), which originate from the two distinct twisted angles, were directly observed. Our results indicate that the TTG could be treated as a subtle superposition of two different TBG, in both topographic and electronic structures. Moreover, the split of the VHSs was also observed by using high-resolution STS measurements, suggesting that electron-electron interactions play an important role in affecting the electronic properties of graphene systems with low-energy VHSs[23].

In our experiment, both the studied graphene bilayer and trilayer were synthesized by a typical low pressure chemical vapor deposition (LPCVD) method on palladium (Pd) foils (See more growth details, Figs. S1-S3 in Supplementary Methods)[38]. We frequently observed twist between adjacent layers in our synthesized samples. There are many reasons, such as the wrinkles, domains, or grain boundaries induced by the multi crystal faces[35], the various crystalline orientations of Pd substrate[39], the thermodynamics and kinetics factors (such as the growth temperature, flow rates of gases and the cooling rate)[40,41], that are likely to break the stable Bernal-stacked mode and introduce mutual misorientations into the graphene layers (See more details in Figure S4 and S5)[38]. Therefore, it is easy to find stacking faults in the synthesized multilayer graphene[2,35,42,43] and the obtained system provides us a fantastic platform to investigate the effects of the twist on the topographic structures and electronic properties of multilayer graphene.

Figure 1(a) shows a typical STM topography of the TTG, which can be undoubtedly identified by the "double-moiré" superlattices. Fast Fourier transforms (FFT) analysis of the STM image in Fig. 1(a) (see the top left inset of Figure 1(a)) shows two sets of hexagonal reciprocal-lattice spots, further confirming the existence of two distinct moiré superlattices in the system. The top right inset of Fig. 1(a) shows a well-defined honeycomb-like hexagonal lattice of the topmost graphene sheet, indicating that there is no obvious distortion of the graphene lattice. To further characterize the structure, we measured the same region of the TTG with different sample biases. Figure 1(b) and 1(c) are two typical images recorded at two different biases, -700 mV and 200 mV, respectively, which unveil two moiré patterns with different periods in the same region (see Figures S6 and S7 for more images scanned at different bias)[38]. The two different moiré patterns are generated by the twist between adjacent layers of the TTG and the insets in Figure 1(b) and 1(c) show the corresponding FFT images. The periods $D$ of the moiré patterns are measured as $D \approx 5.01$ nm and $D \approx 6.70$ nm, respectively, then the corresponding twist angles $\theta$ are estimated to be 2.81° and 2.10° according to $D = a/[2\sin(\theta/2)]$, where a ~ 0.246 nm is the lattice constant of graphene.

In the STM measurements, the images recorded at higher sample bias reflect more superficial information of the sample[44]. Therefore, the 2.81° moiré pattern obtained at the higher scanning bias is generated by the misorientation between the top two graphene layers, and the 2.10° moiré superlattice is induced by the twist between the bottom two layers of the TTG. Additionally, the period of the large moiré pattern in Figure 1(a) is measured to be $D \approx 20.1$ nm, therefore the twist angle is estimated to be $\theta \sim 0.70°$, which nicely matches the difference between 2.81° and 2.10°. In Fig. 1(d-f), we simulate the structure of the TTG according to our experimental result. With considering a moiré pattern generated by a twist angle 2.81° between the top two layers and a moiré pattern induced by a rotation angle 2.10° between the second and third layers, our simulated result is shown in Fig. 1(d). Obviously, it reproduces the main feature of STM image shown in Fig. 1(a), further confirming the structure of the studied TTG.

The twist in the TTG not only results in the complicated moiré patterns, but also separates Dirac cones of the three graphene layers in reciprocal space, as schematically shown in Fig. 2(a). Interactions between each couple of the separated Dirac cones of two adjacent layers generate two saddle points appearing along the intersection of cones and thus result in two VHSs with energy separation of $\Delta E_{VHS}$, as shown in Fig. 2(b). Therefore, it is expected to observe two sets of the VHSs, i.e., four peaks in density-of-state (DOS), in the TTG. To explore the electronic structure of the TTG, which has not been experimentally studied before, we carried out STS measurement on the system. Figure 2(c) shows a representative STS spectrum of the TTG, which exhibits four pronounced peaks in the low energy zone. The four peaks are attributed to the two sets of the VHSs of the TTG and the energy separations of the two sets of the VHSs $\Delta E_{VHS}^1$ and $\Delta E_{VHS}^2$ are measured to be about 179.9 meV and 64.9 meV, respectively. Our simulation of two TBG with two different twist angles (2.81° and 2.10°), as shown in Fig. 2(d), reproduces well the positions of the observed VHSs in the TTG. The calculated local DOS (LDOS) of the TTG by simply adding the LDOS of the two TBG together, as shown in Fig. 2(e), also captures the main feature of our STS spectrum (See more details of theoretical calculation in Supplementary Methods)[38]. Such a result indicates that the electronic structure of the TTG could be simply regarded as a combination of that of two different TBG.

Recent experiments demonstrated that it is possible to realize novel correlated states in TBG with a small twist angle when its low-energy VHSs are quite close to the Fermi level[4,23]. In our experiment, the lowest VHS of the TTG is only about 20 meV away from the Fermi level, therefore, it is reasonable to expect that electron-electron interactions may play an important role in affecting the electronic properties of the studied TTG. To further explore the low-energy electronic properties of the TTG, we carried out high-resolution STS measurements (See more details of high-resolution STS in Supplementary Methods and more STS spectra in Figure S8, S9)[38]. Figure 3(a) shows a typical high-resolution STS spectrum with a spectroscopic resolution of about 1 meV. Compared with the spectrum in Fig. 2(a), where the energy resolution is about 10 meV, it is obviously to find out that each of the four VHSs splits into two peaks (marked by red and blue arrows) with energy separation of ~$(12\pm5)$ meV. The intensity of the split exhibits a slight spatial dependence on the moiré-induced modulation, as shown in Fig. 3(b). Very recently, the split of the VHSs was also observed in a slightly TBG and was treated as an evidence of strong electron-electron interactions induced by the low-energy VHSs[23]. In such a case, symmetry breaking of electronic states around the VHSs was accompanied with the split of the VHSs. Therefore, we measured the electronic states of the TTG in different energies by differential conductance maps (STS maps), which are widely used to directly reflect the spatial distribution of the LDOS at the measured energy. Figures 3(d)-3(h) show the measured LDOS distributions of the TTG in different energies in the same region as shown in Fig. 3(c). At energies far away from the VHSs, the distribution of the LDOS (Fig. 3(d)) reveals the same period and circular symmetry of the moiré pattern but exhibits the inverted contrast comparing with that shown in the STM image (Fig. 3(c)), in consistent with that observed in TBG in previous works[1,16,21,22]. This result indicates the presence of the moiré potential in the TTG due to the twist-induced moiré pattern, which is similar as that observed in the TBG. At energies around the VHSs, the observed features, as shown in Figs. 3(e)-3(h), are quite different. The distribution of the LDOS suffers an apparently spatial distortion, and both the symmetry and period of the moiré pattern are completely disappeared (See more STS maps in Figure S10)[38]. These results suggest that electron-electron interactions are strongly enhanced and break the symmetry of electronic states around the low-energy VHSs of the TTG.

Since the electronic structure of the TTG could be simply regarded as a combination of that of two different TBG, therefore, it is reasonable to expect that the split of the

VHSs should also be observed in TBG with similar twist angles. In Fig. 4, we show STM images and high-resolution STS measurements of two TBG systems with twist angles of 2.01 ° and 1.50 °. We observed the split of the VHSs in both the samples, as shown in Figs. 4(c) and 4(d). The tunneling spectra maps in Figs. 4(e) and 4(f) reveal the spatial evolution of the split along the moiré pattern shown in the right panels. It is apparent that there is a slight modulation of the intensity of the split along the moiré pattern, which is similar as that observed in the TTG (Fig. 3(b)). The above results indicate that electron-electron interactions may be quite important in all the graphene systems with low-energy VHSs.

In conclusion, we systematically studied the topographic structures and low-energy electronic properties of the TTG. Our results demonstrated that both the topography and electronic properties of the TTG could be simply considered as a combination of that of two different TBG. The observed split of the VHSs and symmetry breaking of electronic states around the VHSs in both the TBG and TTG indicates that electron-electron interactions may be quite important in all the graphene systems with low-energy VHSs.

**Acknowledgements**

This work was supported by the National Natural Science Foundation of China (Grant Nos. 11674029, 11422430, 11374035), the National Basic Research Program of China (Grants Nos. 2014CB920903, 2013CBA01603), the program for New Century Excellent Talents in University of the Ministry of Education of China (Grant No.

NCET-13-0054). L.H. also acknowledges support from the National Program for Support of Top-notch Young Professionals and support from "the Fundamental Research Funds for the Central Universities".

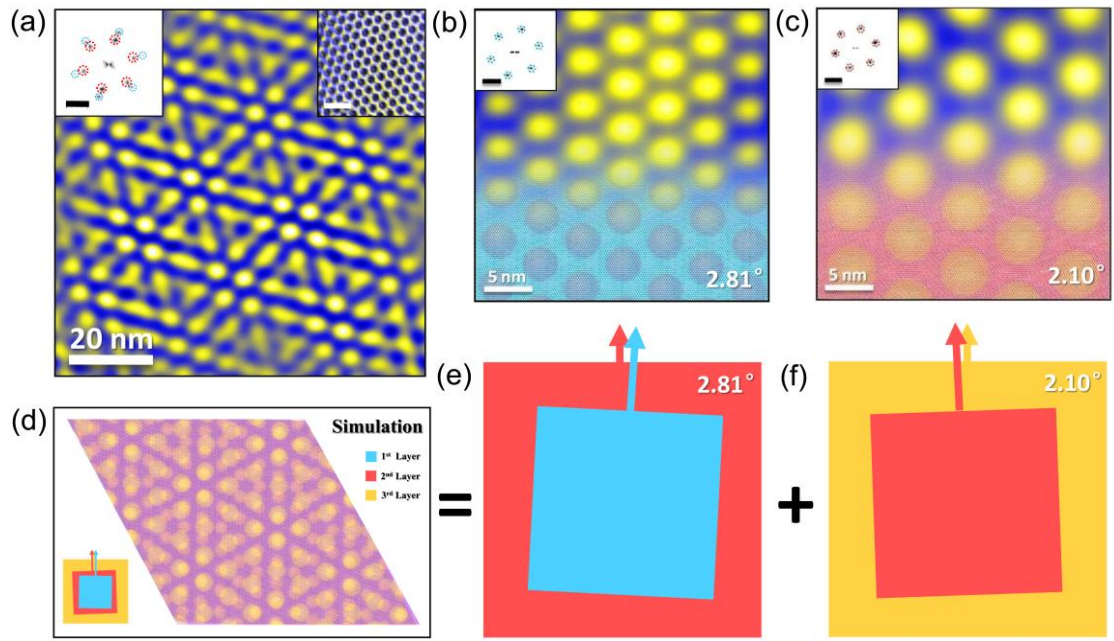

**FIG.1.** (a) A typical STM topography of the TTG (sample bias $V_s$ = -0.6 V, tunnel current $I$ = 320 pA). The top left inset illustrates the Fast Fourier transforms image of the two series of moiré patterns of the TTG [the scale bar is 0.16 (1/nm)]. The top right inset shows a 2 nm × 2 nm STM image of the TTG with perfect honeycomb lattice. (b)

and (c) 40 × 40 nm² STM topographic images of the TTG in the same region measured at different sample biases. The bottom of panels (b) and (c) shows the simulated images of TBG with twist angles of 2.81° and 2.10°, respectively. The insets show the corresponding FFT image [the scale bars in (b) and (c) are 0.17 (1/nm) and 0.15 (1/nm), respectively]. (d-f)The simulation of the TTG with 1$^{st}$ layer (Blue), 2$^{nd}$ layer (Red) and 3$^{rd}$ layer (Orange) as the legend shown (the left bottom inset of panel d is the sketch of the TTG).

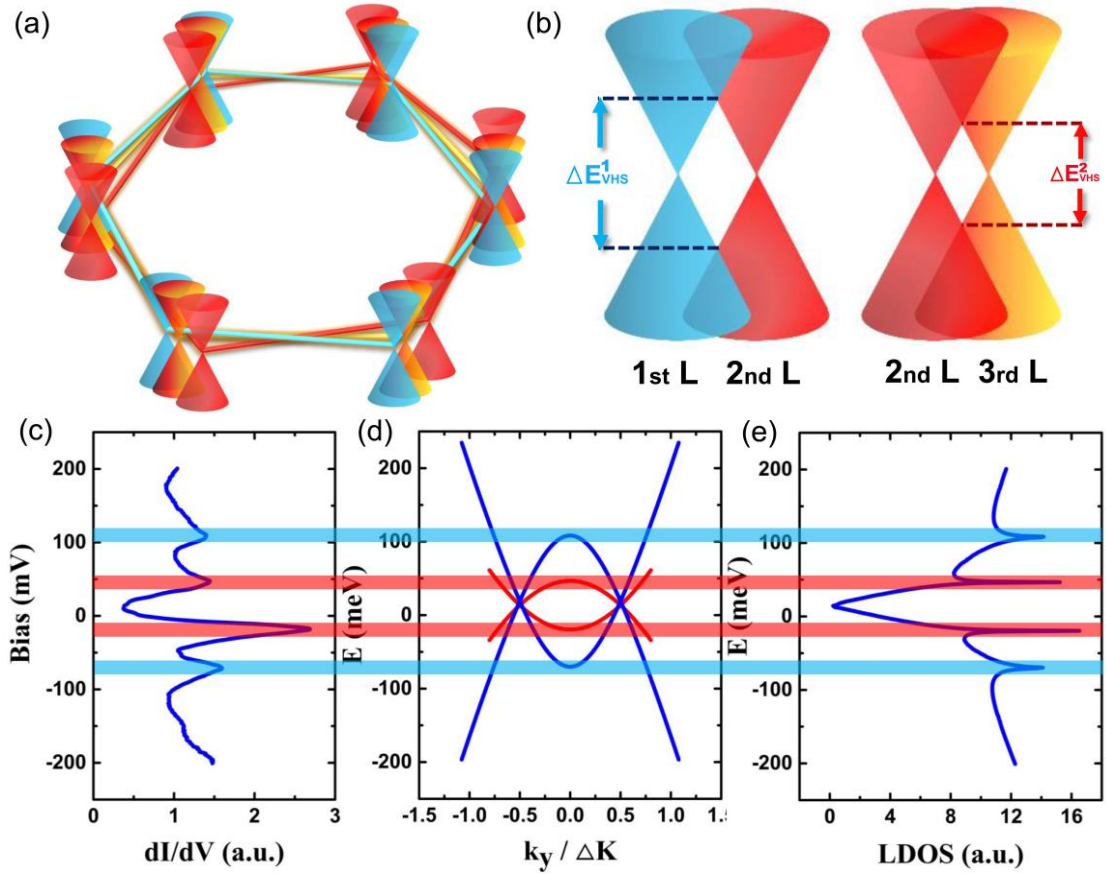

**FIG.2.** (a) The schematic Dirac cones of the three graphene layers of the TTG in the reciprocal space. (b) Zoom-in image of the separated Dirac cones of the three graphene layers in (a). Interactions are expected to generate two VHSs between adjacent layers. Therefore, it is expected to observe four VHSs in the TTG. (c) A typical STS spectrum of the TTG. The two sets of VHSs are marked with red and blue lines. (d) The low-energy band structures of two TBG with rotation angles 2.10°(red) and 2.81°(blue). The main parameters in the simulation are $v_F \sim 1 \times 10^6$ m/s, $t_\theta \sim 300$ meV for the 2.10° system and $v_F \sim 1 \times 10^6$ m/s, $t_\theta \sim 180$ meV for the 2.81° system. (e) The simulated LDOS of the TTG.

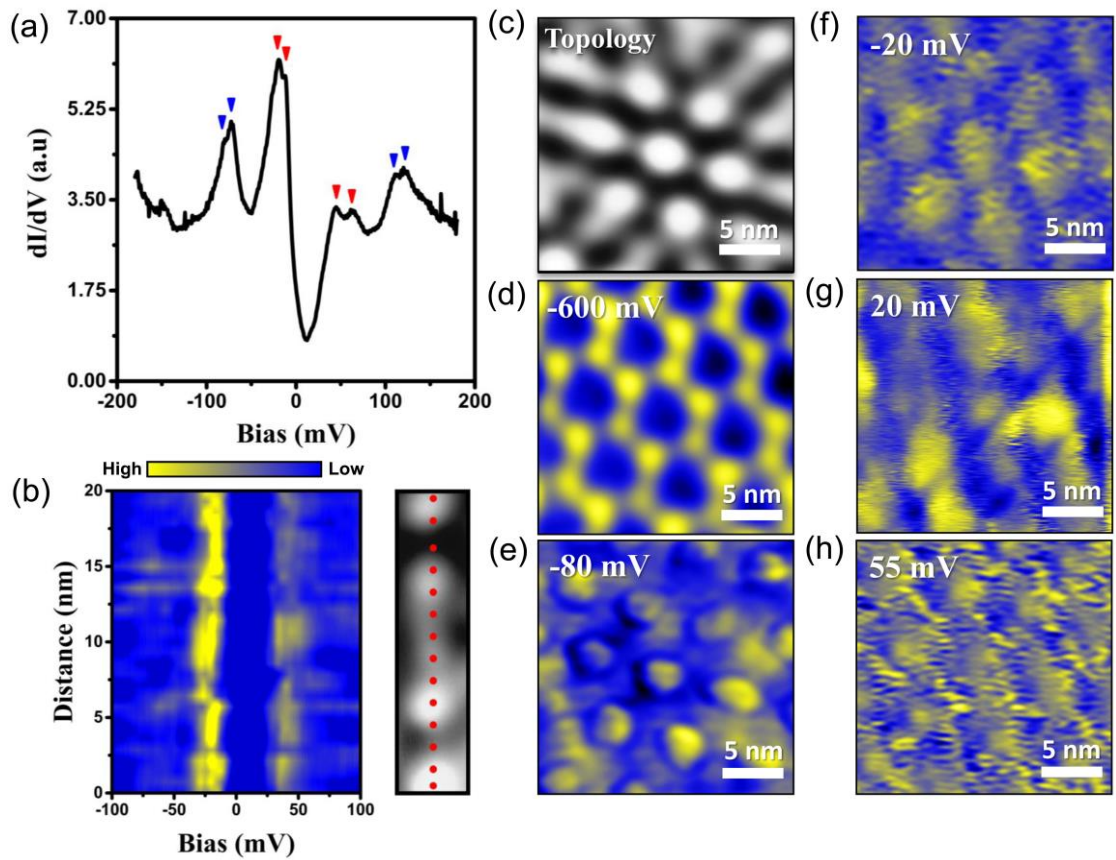

**FIG.3.** (a) A typical high-resolution STS spectrum of the TTG. Each of the four VHSs splits into two peaks. (b) Spatial resolved STS spectra measured along the red dots in right panel. (c) 20 × 20 nm² STM topographic image (sample bias $V_s$ = -0.6 V, tunnel current $I$ = 280 pA). (d)-(h) STS maps measured in the same region of (c) at different sample biases, $V_s$ = -600 mV (d), -80 mV (e), −20 mV (f), 20 mV(g), 55 mV(h).

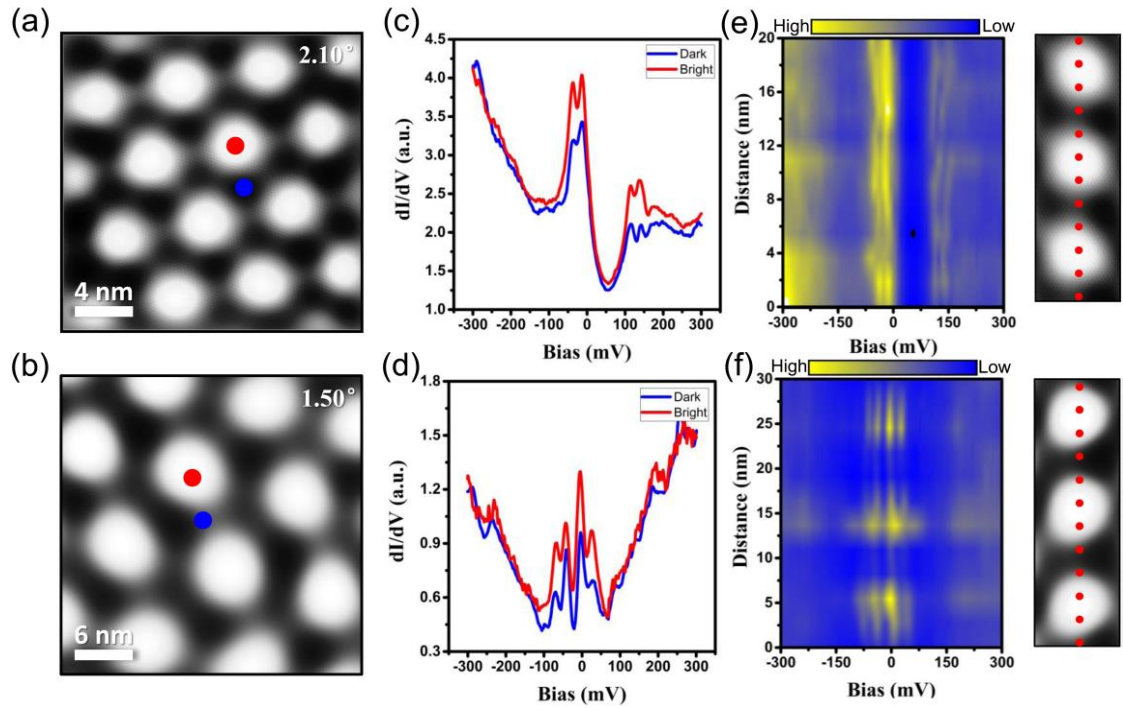

**FIG.4.** (a) and (b) STM topographic images of TBG with $\theta = 2.01°$ and $\theta = 1.50°$, respectively. (c) and (d) Typical high-resolution STS spectra measured in the marked blue and red points in (a) and (b), respectively. (e) and (f) Spatial resolved high-resolution STS spectra measured along the red dots in right panel for the 2.01° and 1.50° TBGs, respectively.